\documentclass[runningheads]{llncs}
\usepackage{graphicx}
\usepackage{cite}
\usepackage[bookmarks=false,hidelinks]{hyperref}
\hypersetup{
    colorlinks=true,
    linkcolor=blue,     % color of internal links
    citecolor=blue,     % color of citation links
    filecolor=blue,     % color of file links
    urlcolor=blue       % color of external links
}
\usepackage{booktabs} % For better-looking tables

\usepackage{amsfonts}
\usepackage{amsmath}
\DeclareMathOperator*{\argmax}{arg\,max}

\begin{document}
\title{Multi-Track MusicLDM: Towards Versatile Music Generation with Latent Diffusion Model}  %\thanks{Supported by organization x.}}
\titlerunning{MT-MusicLDM}

\author{Tornike Karchkhadze\inst{1} \and
Mohammad Rasool Izadi\inst{2} \and
Ke Chen\inst{1} \and
Gerard Assayag\inst{3}\and
Shlomo Dubnov\inst{1}}
\authorrunning{T. Karchkhadze et al.}
\institute{UC San Diego, CA 92093, USA \and
Bose Corp., Framingham, MA 01701, USA \and
Institute for Research and Coordination in Acoustics and Music (IRCAM), Paris, France \\
\email{\{tkarchkhadze, kchen, sdubnov\}@ucsd.edu, Russell\_Izadi@bose.com, Gerard.Assayag@ircam.fr}}

\maketitle            % typeset the header of the contribution
\begin{abstract}
Diffusion models have shown promising results in cross-modal generation tasks involving audio and music, such as text-to-sound and text-to-music generation. These text-controlled music generation models typically focus on generating music by capturing global musical attributes like genre and mood. However, music composition is a complex, multilayered task that often involves musical arrangement as an integral part of the process. This process involves composing each instrument to align with existing ones in terms of beat, dynamics, harmony, and melody, requiring greater precision and control over tracks than text prompts usually provide. In this work, we address these challenges by extending the MusicLDM—a latent diffusion model for music—into a multi-track generative model. By learning the joint probability of tracks sharing a context, our model is capable of generating music across several tracks that correspond well to each other, either conditionally or unconditionally. Additionally, our model is capable of arrangement generation, where the model can generate any subset of tracks given the others (e.g., generating a piano track complementing given bass and drum tracks). We compared our model with existing multi-track generative model and demonstrated that our model achieves considerable improvements across objective metrics, for both total and arrangement generation tasks. Sound examples can be found at \href{https://mt-musicldm.github.io}{https://mt-musicldm.github.io}

\end{abstract}

\section{Introduction}\label{sec:introduction}

In recent years, diffusion models~\cite{ho2020denoising} have demonstrated their ability to learn complex distributions, rendering them well-suited for data types such as raw audio. These advancements have significantly impacted the domains of speech and general audio generation~\cite{pmlr-v202-liu23f, liu2023audioldm, Yang2022, huang2023makeanaudio, kong2021diffwave, zhu2023edmsound, yuan2023textdriven, karchkhadze2024latent, luo2023difffoley, ijcai2022p577, zhang2023survey, 9746901}, as well as the generation of music directly within the audio domain~\cite{huang2023noise2music, Forsgren_Martiros_2022, chen2023musicldm, melechovsky2024mustango, wu2023music, schneider2023mousai, novack2024ditto}. User-controlled neural audio synthesis, particularly in music generation, has the potential to revolutionize music industry by providing musicians with tools for quick compositional prototyping, speeding up the creative process. Additionally, these tools contribute to the democratization of the field by allowing amateur musicians to leverage generated pieces to compose without extensive knowledge of instruments, music theory and years of musical education.

Most generative music models operate by conditioning music on high-level ideas expressed as text, such as genre and mood, leading to text-to-music (TTM) task. However, TTM paradigms have conceptual problems. One challenge is that music, as an abstract entity, is generally difficult to describe with words. Another issue is that text is not an effective medium to convey time-dependent musical attributes, which are crucial for musical expression. Furthermore, music is a multilayered art in which many tracks of instruments simultaneously play their unique roles while being in correspondence with each other on lower-level attributes like notes, timbre, dynamics, harmony, and rhythm.  The essence of musical composition often boils down to arrangement—structuring the piece, orchestrating interactions, and determining the overall sonic character by distributing texture among different instruments or voices, a complexity that is challenging to convey through text.

To bridge the conceptual gaps in current music generation models, we introduce the Multi-Track MusicLDM, a diffusion-based model that generates coherent music in multiple tracks or stems (terms we use interchangeably), ensuring they correspond and collectively create a unified musical piece. To achieve this, we utilized the MusicLDM~\cite{chen2023musicldm} model, which is an adaptation of AudioLDM~\cite{pmlr-v202-liu23f} for music, and transformed it into a multi-track audio diffusion model. Taking a multi-track inspiration from the recent work MSDM~\cite{mariani2024multisource} and operating on the latent space,  our model learns a joint probability distribution for tracks that share a contextual structure and generates music mixtures in separate tracks, a process referred to as total generation. Leveraging the text and music conditioning capabilities of the CLAP~\cite{wu2023largescale} component, our model provides options for additional conditioning. Audio conditioning can be applied by using an existing reference track processed through CLAP's audio branch to influence the generation. This track can guide the overall character and content of the generated music, effectively enabling a transfer that preserves the essence while allowing creative deviations. Similarly, text conditioning can influence the genre, mood and overall character of the generated tracks. Additionally, by employing a well-known method from diffusion model, inpainting, our model is capable of imputation of tracks or generating any subset of tracks given others. We refer to this process as arrangement generation. Arrangement generation enables, for example, the creation of specific musical parts, like piano or guitar, to accompany existing bass and drum tracks. The conditioning mentioned above can be used in combination with arrangement generation, giving our model an additional edge in creative endeavors. By designing desired instrument combinations and using audio and text conditioning, users have the flexibility to generate specific arrangements or full musical pieces, tailoring the model to their compositional needs.

In our experiments, we demonstrate that our model can generate realistic music across various scenarios: total track-by-track music generation, conditional generations, and arrangement generation with any combination of stems. Furthermore, we compared our model with the existing open source multi-track generative model, MSDM, and demonstrated that our model, trained on the same dataset, achieves considerable improvements in the Fréchet Audio Distance (FAD)~\cite{kilgour2019frechet} score compared to the baseline. 

% As part of our commitment to reproducibility and open science, the code and checkpoints of this study will be made available on GitHub upon acceptance of this paper.
\setcounter{footnote}{0}
As part of our commitment to reproducibility and open science, the code and checkpoints of this study are publicly available \footnote{\href{https://github.com/karchkha/mt-musicldm}{https://github.com/karchkha/mt-musicldm}}.

\begin{figure*}[t]
  \centering
  \includegraphics[width=.98\linewidth]{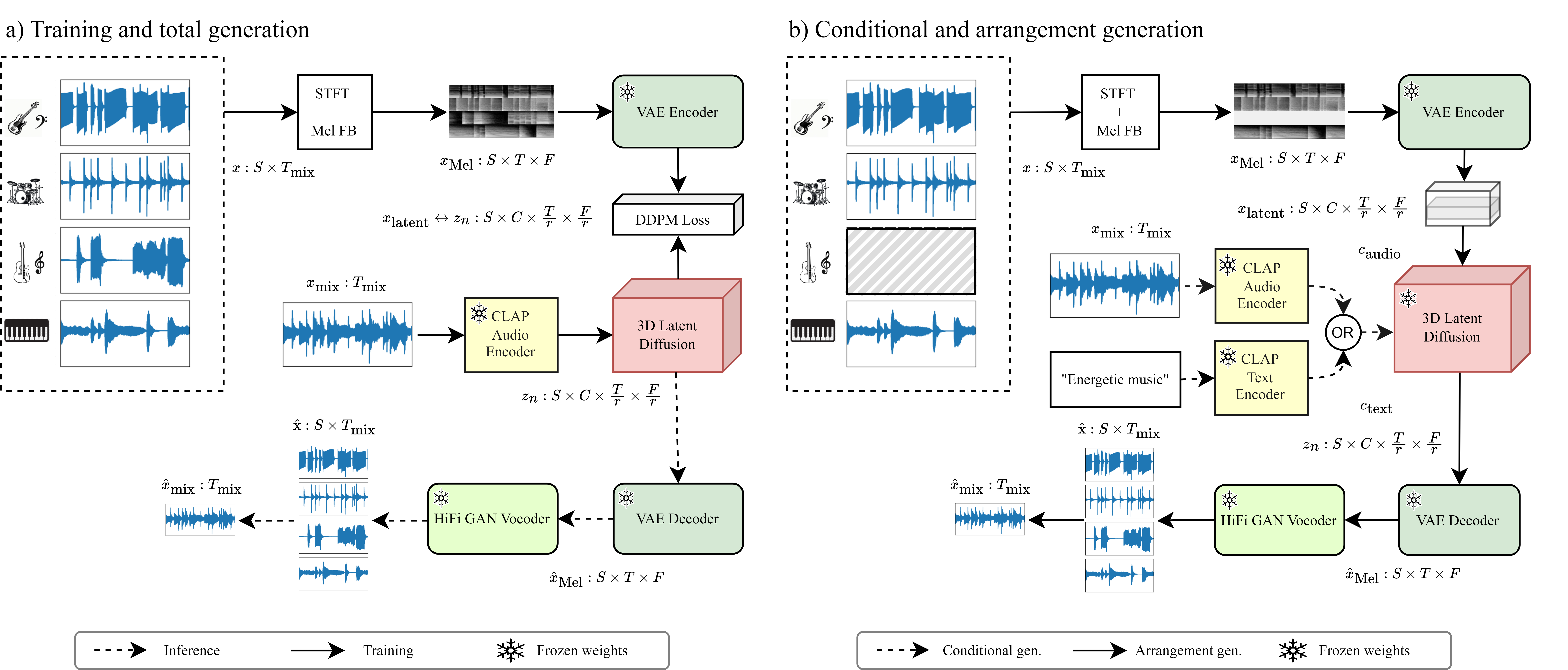}
  \setlength{\abovecaptionskip}{-1pt}
  \caption{Multi-Track MusicLDM system overview: a) During training, our model processes audio stems that are converted into Mel-spectrograms. A VAE encoder then compresses these spectrograms into a 3D latent space, where LDM operates. For conditional training, model takes an audio mixture as conditioning thought a CLAP input. During inference, the model generates audio stems unconditionally, where the generated latent vector is first up-sampled back to a Mel-spectrogram by VAE decoder and then converted into audio via HiFi-GAN. b) For arrangement generation, our model takes as an input a set of given tracks to add (inpaint) the missing ones. In conditional generation, it takes text or a reference music track as input though CLAP and uses it to condition the LDM.}
  % \vspace{-9pt}
  \label{fig1}
\end{figure*}

\section{Related Work} 
\label{section_Background}

% \subsection{MusicLDM/AudioLDM overview}

\textbf{Music Generation.} Our system, Multi-Track MusicLDM (MT-MusicLDM), depicted in Figure~\ref{fig1}, is an extended version of the MusicLDM model capable of learning and generating multiple simultaneous music stems. MusicLDM shares its architecture with AudioLDM, which in turn is based on a Latent Diffusion Model (LDM)~\cite{ho2020denoising} framework with a cascaded model structure. The system comprises a text-audio encoder, an LDM generator, a Variational Autoencoder (VAE)~\cite{kingma2022autoencoding}, and a vocoder. The role of text and audio encoding is played by CLAP~\cite{wu2023largescale}, serving as the system's encoder and mapping audio and text into a shared embedding space for conditioning. This is followed by an LDM that acts as the main generator, operating in the latent space (instead of directly on audio or spectrograms), allowing for training and inference on limited computational resources while retaining generation quality. This latent space is achieved by a  VAE that manages the data dimensionality, pre-trained to compress and reconstruct Mel-spectrograms into latent representations. Finally, a HiFi-GAN~\cite{kong2020hifi} vocoder synthesizes the audio output from the generated Mel-spectrograms. 

\textbf{Arrangement generation.}  A commercial product JEN-1 Composer \cite{yao2023jen1} introduced generating musical arrangements in audio domain using a latent diffusion-based music generation system. However, unlike our approach, JEN-1 Composer is not multi-track and uses a latent space obtained directly from audio with its own pre-trained autoencoder model for audio reconstruction, following its predecessor, the Jen-1 architecture proposed in \cite{li2023jen}. Additionally, the dataset used is also different from ours, as they used their own private dataset, which is larger and of higher quality than the publicly available Slakh2100~\cite{Slakh} dataset that we used. Unfortunately, as it is a commercial product, they do not share code, datasets, or provide a free API for audio generation, making direct comparison to our work impossible. Other recently published works have explored music-to-music and arrangement-like music generation. STEMGEN \cite{stemgen2024} presented an alternative paradigm for music generation by introducing a model that can listen and respond to musical context. Different from ours, STEMGEN uses a transformer language model-like architecture on hierarchical discrete representations from VQ-VAEs \cite{van2017neural} to model musical tracks. Diff-A-Riff~\cite{nistal2024diffariff} presented a non-multi track approach to generating instrument accompaniments using diffusion models and incorporates control mechanisms like both text-driven and audio-driven CLAP conditioning. The most similar work to ours is \cite{xu2024multisourcemusic}, which was published during the review process of our paper. This study also explored multi-track music generation using the same Slakh2100 dataset and reported results that are very similar to ours.

Some other works approach the arrangement task with a one-to-many or many-to-one paradigm. For example, \cite{pasini2024bass} generates bass accompaniment using audio autoencoders. On the other hand, SingSong \cite{donahue2023singsong}, utilizing an architecture similar to AudioLM \cite{borsos2023audiolm}, proposed a method for generating instrumental music to accompany input vocals and demonstrated promising results. However, these works primarily concentrate on one instrument, offering limited versatility. Another work that serves as our baseline, MSDM \cite{mariani2024multisource}, introduced a novel diffusion-based, multi-source generative framework trained via denoising score-matching \cite{song2019, song2021}. MSDM is capable of synthesizing music, creating arrangements, and separating musical sources while operating in the raw waveform domain. While MSDM is notable for its flexibility, it exhibits limitations in audio quality and musical coherence. Interestingly, although our model surpasses MSDM in most tasks, it struggles with source separation—a limitation we attribute to its operation within a latent space where mixtures and stems do not maintain a linear relationship. This insight is prompting our future research in this direction.

Arrangement generation was extensively studied in the symbolic music domain. Using the audio-to-MIDI generation paradigm, these works typically focus on generating harmonization for a given melody, such as in \cite{Paiement2006ProbabilisticMH, Yeh2020}, or extracting pitch information from input vocals and generating chords suitable for the melody, as seen in the commercial product Microsoft Songsmith, inspired by \cite{simon2008mysong}. Other significant contributions include the works by \cite{lattner2019high} and \cite{grachten2020bassnet}, which explored generating kick drum and bass accompaniments, respectively, in the MIDI domain. The multi-track MIDI music generation paradigm was also employed in systems like MuseGAN \cite{MusGan2018}, MIDI-Sandwich \cite{MIDISandwich19}, Multitrack Music Machine \cite{MMM2020}, and MTMG \cite{MTMG20}. Additionally, MIDI accompaniment generation based on audio conditioning was suggested in \cite{Deruty2022}.

% additonal realted work comed form symbolic music domain using audio to MIDI generation paradigm. there works typically focus on generating harmonization for a given melody such as~\cite{Paiement2006ProbabilisticMH,Yeh2020}. Other notable works is Microsoft Songsmith, inspired by~\cite{simon2008mysong}, which exemplifies this approach by extracting pitch information from input vocals and generating chords suitable for the melody. Other body of work, such as \cite{lattner2019high}  and \cite{grachten2020bassnet}, have explored generating kick drum and bass accompaniments, respectively in MIDI domain. Multi-track MIDI music generation paradigm was also emplyed  in such wworks systems like MuseGAN\cite{MusGan2018}, MIDI-Sandwich\cite{MIDISandwich19}, Multitrack Music Machine \cite{MMM2020} and MTMG \cite{MTMG20}. accompaniment generation was aslo suggested based on audio conditioning in \cite{Deruty2022}.

\section{Method}

We introduce MT-MusicLDM, depicted in Figure~\ref{fig1},  as an integration of MusicLDM and MSDM. In section 1, we expand the LDM latent space to adopt MusicLDM for multi-stem music generation. This expansion enhances our control and versatility over the music generation process. Next, in section 2, we create musical arrangements for a given subset of tracks. Section 3 explores the use of CLAP encoders to control the style of music with text and musical prompts.

\subsection{Multi-Track LDM}

% In our work, we handle multiple tracks or stems (terms we use interchangeably) simultaneously. 

Following AudioLDM and MusicLDM work, we employ denoising diffusion probabilistic models (DDPMs)~\cite{ho2020denoising, Sohl-DicksteinW15} for audio/music generation. DDPMs belong to a category of latent generative variable models. Given two mappings between a time-domain sample \(x\) and its corresponding latent space representation \( x_\text{latent}\), i.e. \(x \leftrightarrow x_{\text{latent}} \), the generation problem is to model \(q(x_{\text{latent}})\) instead of \(q(x)\). We first describe the mappings to and from the latent-space for multi-track music and then explain the training and inference of the LDM.

% \textbf{MT-LDM}
Let \(x_{\text{mix}}\) be a mixture composed of \(S\) stems \(x_s\) with a duration of \(T_{\text{mix}}\) for \(s \in \{1, \dots, S\}\) such that \(x_{\text{mix}} = \sum_{s=1}^S x_s\). Also, denote the stack of stems as \(x\) with dimensions \(S \times T_{\text{mix}}\).
As shown in Figure~\ref{fig1}, the individual stack of stems \(x\) is transformed into a Mel-spectrogram \(x_{\text{Mel}}\), using short-time Fourier transform (STFT) and Mel operations, with dimensions \(S \times T \times F\) where \(T\) and \(F\) show the time and frequency dimensions of Mel-spectrogram, respectively. Subsequently, the VAE encoder translates \(x_{\text{Mel}}\) into a latent, compressed representation \(x_{\text{latent}}\) with dimensions \(S \times C \times \frac{T}{r} \times \frac{F}{r}\), where \(r\) indicates the VAE's compression ratio, and \(C\) denotes the number of channels in the latent space. Following this, the VAE decoder converts the latent, compressed representation, \(x_{\text{latent}}\), to the Mel-spectrogram domain as \( \hat{x}_{\text{Mel}}\), which is then transformed to the time-domain \( \hat{x} \) by the HiFi-GAN vocoder. Therefore, the time-domain to latent space mapping, \( x \rightarrow x_{\text{latent}}\), is composed of STFT, Mel, and VAE encoder, while the reverse mapping \(x_{\text{latent}} \rightarrow x\), employs VAE decoder and HiFi-GAN vocoder.

The generative model for \( x_{\text{latent}}\) via the diffusion process is defined as \( p_{\theta}(z_0) = \int p_{\theta}(z_{0:N}) dx_{1:N} \) where \( z_0 = x_{\text{latent}}\) is the latent representation and $\theta$ corresponds to the parameters of the LDM model. Within the DDPM framework, the LDM generator operates in the latent space to generate a latent representation \(z_0 \sim q(z_0)\), either conditionally or unconditionally, from Gaussian noise \(z_N \sim \mathcal{N}(0, \sigma_N^2 I)\). The variable \( n \) in \( z_n \), where \( n \in [1, \ldots, N] \), represents the step number in the diffusion model's forward or reverse process, and \(N\) denotes the number of total steps. The forward pass gradually introduces Gaussian noise, \(\epsilon \sim \mathcal{N}(0, I)\), to \(z_0 = x_{\text{latent}}\), i.e. \( z_n = z_0 + \sigma_n \epsilon\) ultimately resulting in isotropic Gaussian noise \(z_N\) with a distribution \(\mathcal{N}(0, I)\) over \(N\) steps. Conversely, the reverse process aims to eliminate noise by estimating the injected noise, \(\epsilon\), at every step and obtaining \(z_{n-1}\) from \(z_n\), thereby incrementally reconstructing \(z_0 = \hat{x}_{\text{latent}} \). The DDPM parameterizes the reverse Gaussian distribution \( p(z_{n-1}|z_n)\) with a neural network \( \epsilon_{\theta}(z_n, n) \):
\begin{gather}
    p(z_{n-1}|z_n) \!=\! \mathcal{N}( z_{n-1} | \mu_{\theta}(z_n, n), \sigma_n^2) \label{eq:cond} \\
    \mu_{\theta}(z_n, n) \!=\! ((\sigma_n^2 - \sigma_{n-1}^2) \epsilon_{\theta}(z_n, n) + \sigma_{n-1}^2 z_n ) / {\sigma_n^2}.
\end{gather}
Optimizing the evidence lower bound on the log-likelihood \( q(z_0)\) simplifies to minimizing the mean squared error between the predicted noise \(\epsilon_{\theta}\) and the Gaussian noise \(\epsilon\), at each step, as follows:
\begin{equation}
L(\theta) = \mathbb{E}_{z_0, \epsilon, n} \|\epsilon - \epsilon_{\theta}(z_n,n, [c_{\text{cond}}]) \|^2
\label{eq:loss}
\end{equation}
where \([c_{\text{cond}}]\) denotes the optional use of conditioning, which we will touch upon in more detail later in the section. Therefore, having the time-domain mapping, \(x \rightarrow x_{\text{latent}} \), one can estimate the LDM \( \theta \) by minimizing the loss function \( L(\theta)\), presented in Eq.~\ref{eq:loss}. For inference, the generation begins with the Gaussian noise prior \( z_N \), followed by an iterative backward sampling process from \(p(z_{n-1}|z_n)\) for each \(n \in \{N, \dots, 1\}\), as outlined in Eq.~\ref{eq:cond}. The final step involves mapping the generated latent representation, \( \hat{x}_{\text{latent}} = z_0 \), to the time-domain using \( x_{\text{latent}} \rightarrow x \). To form the mixture of tracks, one can simply add all the rows of \(\hat{x}\), i.e. \( \hat{x}_{\text{mix}} = \sum_{s=1}^S \hat{x}_s \), with each \(\hat{x}_s \) being the \(s\)-th row of \(\hat{x}\).

% Fast LDM
For high-quality reconstructions using DDPM, a large number of steps, typically \(N = 1000\), are traditionally necessary. However, to streamline the process and reduce computational demands, denoising diffusion implicit models (DDIM)~\cite{song2022denoising} offer a compelling alternative. In this study, we utilize the DDIM protocol, which allows for a substantial reduction in the number of required steps, down to approximately \(N = 200\) during inference, while still preserving the generative quality.

% architecture
As Audio/MusicLDM frameworks for audio and music generation are heavily inspired by methods originally used in image generation, their network architecture is borrowed from this domain. Like in image generation, a large UNet~\cite{RonnebergerFB15} architecture is a common choice for diffusion models in audio and music domain. The UNet architecture consists of two symmetrical halves: an encoder and a decoder, both enhanced with skip connections that bridge corresponding layers. To accommodate \(z_n\) with an additional dimension compared to typical image or single-channel audio representations, \(S \times C \times \frac{T}{r} \times \frac{F}{r}\) vs \(\times C \times \frac{T}{r} \times \frac{F}{r}\) in Audio/MusicLDM, we extend the UNet architecture by using 3D convolutional operations. We effectively enhance the operational dimensionality of UNet by interpreting the channel dimension of \(z_n\) as an additional spatial dimension. This adjustment leads to the stems’ dimension now serving as the channel dimension.

\subsection{Conditional Generation}
To have control over the generation \( p_{\theta}(z_0)\), one can introduce some condition \(c\) to the diffusion process, resulting in \( p_{\theta}(z_0|c)\). The conditional diffusion process is defined similarly to the unconditional process \( p_{\theta}(z_{0:N} )\) as follows:
\begin{equation}
p_{\theta}(\mathbf{z}_{0:N} | c) = p(\mathbf{z}_N) \prod_{n=1}^{N} p_{\theta}(\mathbf{z}_{n-1} | \mathbf{z}_n, c).
\label{eq:cond_gen}
\end{equation}
Additionally, to enable more controllable generation, LDM  models often employ classifier-free guidance (CFG)~\cite{ho2022classifierfree}.
CFG is a technique in diffusion models that enhances control over the adherence to conditioning information during inference. 
This is achieved by randomly dropping the conditioning information during training, thereby simultaneously training both conditional and unconditional versions of the LDM model. In the inference time, the strength of the conditioning can be modulated by the CFG weight \(\hat{\epsilon} = w\epsilon_{c} + (1 - w)\epsilon_{u}\), where \( w \) is the guidance scale weight that balances the model's unconditional \( \epsilon_{u} \) and conditional \(\epsilon_{c}\) predictions.

In this study, we employ CLAP to convert text prompts and musical tracks into embeddings, which serve as the basis for conditioning the LDM.
For example, users can specify the type of guitar by conditioning on the CLAP embedding of a reference track.
Additionally, the strength of the conditioning can be adjusted using the CFG weight, increasing their creative options.

\subsection{Arrangement generation}
From a musical perspective, arrangement composition refers to the task of creating plausible musical accompaniments for a particular subset of given tracks. In a broader machine learning literature setting, the task of filling a partially observed the data is commonly referred to as imputation (inpainting in the image domain) and aims to fill out the missing segments of variable. Learning the joint distribution of musical tracks offers us a clear path to explore this task in the latent space.

For a given subset of tracks, \(x_I = \{ x_s | s \in I \} \), arrangement generation task is to find \(x_{\bar{I}} = \{ x_s | s \in \bar{I}\} \), as follows
\begin{equation}
\argmax_{x_{\bar{I}}} p_{\theta}( x_{\bar{I}} | x_I),
\end{equation}
where \(\bar{I} = \{1, \dots, S\} - I\).
Given the LDM, the search for \( x_{\bar{I}} \) happens in the latent space, i.e. \( z_{\bar{I}} \) given \( z_{I} \).
Note that given \(I\), one can find the latent representation of \(x_I\) and \( x_{\bar{I}} \) as follows:
\begin{gather}
x_I \rightarrow m \odot z_0, \\
x_{\bar{I}} \rightarrow (1 - m) \odot z_0, 
\end{gather}
where \( m \) is a binary mask in the latent space with the same dimension as \( z_0 \) and \(m_s = 1 \) if \(s \in I\) and \(m_s = 0 \) otherwise.

The generation of \( x_{\bar{I}} \) starts with sampling a Gaussian noise \(p(z_N) \sim N(0, I)\) as the total generation, but each denoising step in the reverse process if followed by
\begin{equation}
z_{n-1} \leftarrow (1-m) \cdot z_{n-1} + m \cdot z'_{n-1}.
\end{equation}
where \( z'_{n-1} \sim \mathcal{N}(z_{n-1} | z_0 , \sigma^2_{n-1})\) is obtained by adding \(n-1\) noise steps to \(z_0\) through the forward process.

In essence, arrangement generation becomes a generation problem where, at every step, parts of the latent space corresponding to the given tracks are masked and replaced with their noise-added version. This approach compels the model to perform generation under constraints, ensuring that the generated arrangement tracks align well with the given ones.

% ===============================================================================================================

\section{Experimental Setup} \label{section_experimental_setup}

\subsection{Dataset}

Following a similar research path as MSDM and to facilitate direct comparison, we used Slakh2100~\cite{Slakh}, a dataset widely recognized as a benchmark in the domain of music source separation. Although source separation was not our primary focus, the choice of Slakh2100 was motivated by our need for clean and high-quality multi-track audio data for learning multi-track generation. Synthesized from MIDI files using premium virtual instruments, the dataset consists of 2100 individual tracks into subsets of 1500 for training, 375 for validation, and 225 for testing. While the original Slakh compilation offers up to 31 distinct instrumental classes, our experiment and subsequent analyses were limited to \(S=4\) most prevalent classes: Bass, Drums, Guitar, and Piano. These classes were selected due to their dominant presence in the dataset, ensuring a robust and consistent basis for our evaluations.

In our experiments, we performed preprocessing steps on the dataset. We downsampled the audio from its original 22.05 kHz to 16 kHz to align with the specifications of our model. We read audio from the original tracks, creating audio segments of 10.24 seconds with an additional random shift for training samples. To convert the audio clips into a suitable feature representation, we utilized a window length of 1024 and a hop size of 160 samples to generate Mel-spectrograms with dimensions \( F \times T = 64 \times 1024 \). For the creation of mixed audio samples, individual tracks were combined to form mixtures. Differing from the original MusicLDM, we abstained from normalizing the separate tracks or their mixtures to prevent any potential peaking in the audio signals.

\subsection{Model, Training and Evaluation Specifics}

Our parameter configuration closely mirrors that of MusicLDM~\cite{chen2023musicldm}, with only minimal adjustments. We extended the LDM model to accommodate the stem dimension of \(S=4\), transforming it into a 3D LDM model. For our new 3D LDM model, we employed a UNet architecture comprising 2 encoder blocks, a middle block transof, and 2 decoder blocks. We maintained the settings consistent with previous configurations of MusicLDM, with the sole modification being the adaptation to 3D convolutional layers. Additionally, we switched from the "spatial transformer" used by MusicLDM and AudioLDM to a generic attention block transformer, as the former was tailored for 2D, picture-like data, which did not align with our model's 3D data processing requirements.

To obtain a latent representation of Mel-spectrograms, we employed a mono-track VAE from the MusicLDM model, which boasts a compression ratio \( r = 4 \), effectively encoding stacks of Mel-spectrograms of size \( S \times T \times F = 4 \times 1024 \times 64 \) into a 3D LDM latent vector with dimensions \( S \times C \times \frac{T}{r} \times \frac{F}{r} = 4 \times 8 \times 256 \times 16 \), which respectively represent stems, channels, time, and frequency. The components of MusicLDM—including the CLAP encoder, VAE, and the HiFi-GAN vocoder—were taken from the pre-trained, publicly released checkpoint of MusicLDM\footnote{\href{https://github.com/RetroCirce/MusicLDM}{https://github.com/RetroCirce/MusicLDM}}. This checkpoint was trained on an extensive collection of music audio data. However, it is worth mentioning that these components were not trained or fine-tuned for processing separate tracks, which imposes certain limitations on the final audio quality of our model. These components remained unchanged during the training phase of the 3D LDM generator, as illustrated in Figure \ref{fig1}. We utilized pre-trained weights for MusicLDM components from publicly available checkpoints.%\footnote{\href{https://github.com/RetroCirce/MusicLDM}{github.com/RetroCirce/MusicLDM}}.

We trained our model, MT-MusicLDM, with a dropout rate of 0.1 applied during conditional generation, effectively resulting in training both unconditional and conditional models. Training was conducted using the Adam optimizer with a learning rate of \( 3 \times 10^{-5} \) for a duration of up to 1000 epochs. The number of denoising steps for the LDM is set at \( N = 1000 \) during training and reduced to \( N = 200 \) for DDIM sampling during inference. 

We evaluated our models using the Frechet Audio Distance (FAD)~\cite{kilgour2019frechet} metric, a widely recognized benchmark in music quality assessment. This metric was employed across all our experiments, including total generation, arrangement generation, and both audio- and text-conditioned tasks.

%To assess source separation, we utilized the scale-invariant Signal-to-Distortion Ratio improvement (SI-SDRi) metric ~\cite{LeRoux2018SDRH}. 

% \vspace{-0.3cm}
\begin{table}[h]
    \centering
    \caption{FAD score comparison for total music generation tasks. Two variants of our MT-MusicLDM model—unconditional and conditional with audio conditioning—are compared with a baseline, along with a standard MusicLDM model. Note: The asterisk (*) for MusicLDM indicates that evaluations were conducted on different datasets, suggesting that direct comparisons may not fully reflect performance.} 
    \begin{tabular}{lc}
    \toprule
    Model & FAD $\downarrow$\\
    \midrule
    MSDM (Baseline) & 6.55 \\
    MT-MusicLDM ({\textit{Uncond}}) & 1.36 \\
    MT-MusicLDM ({\textit{Audio Cond}}) & \textbf{1.13} \\
    MusicLDM* & 1.68 \\
    \bottomrule
    \end{tabular}

    \label{table:FAD_total_generation}
    \vspace{-7pt} % Adjust or remove based on your layout needs
\end{table}

\section{Experiments and Results} 

\subsection{Total Generation}

We evaluated our MT-MusicLDM model in unconditional mode on the total music generation task using the Slakh test dataset. For this evaluation, we generated audio stems, mixed them to create mixtures, and then calculated the FAD between these generated mixtures and the mixtures from the test set. Given that MSDM was the first and only open source model capable of generating music in individual parts, we selected it as our baseline for comparison. Table ~\ref{table:FAD_total_generation} shows the performance of our model, reported as "MT-MusicLDM ({\textit{Uncond}})", compared to MSDM on the same dataset. We observed that our model significantly outperforms MSDM, with a dramatic reduction in FAD scores from 6.55 to 1.36. This substantial improvement highlights our model's capability to generate high-quality and coherent music audio track-by-track.

For broader context, we also incorporated the benchmark MusicLDM scores from~\cite{chen2023musicldm} in the table. We took the highest-performing variant, "MusicLDM w/ BLM Text-Finetune." This model underwent specialized training with beat-synchronous latent mixup and was further enhanced through fine-tuning on text prompts. We denote MusicLDM with an asterisk in the table to indicate that evaluations were performed on distinct datasets: Audiostock \cite{chen2023musicldm} for MusicLDM and Slakh for our model. Given the significant differences in dataset characteristics these comparisons should be viewed as contextual rather than direct.

\subsection{Audio Conditional Generation}

In the audio-conditioned experiment, we explored the MT-MusicLDM conditioned through CLAP audio branch, focusing on total audio generation with existing audios used for conditioning. We utilized audio stems mixed from the Slakh test dataset as inputs for our model's CLAP encoder and generated audio by conditioning our model with a CFG weight of \( w =2.0\). Subsequently, these generated stems were summed to form new audio mixtures. We then calculated the FAD score between these mixtures and the Slakh test set to evaluate the model's performance in generating coherent audio outputs. In the Table ~\ref{table:FAD_total_generation}, we reported our result as "MT-MusicLDM ({\textit{Audio Cond}})." It is evident that audio conditioning with CLAP noticeably steers generation towards the test set, resulting in further improvement in FAD score. This validates our hypothesis and underscores the potential of our model 
for audio content adaptation with CLAP conditioning, leveraging a user-selected reference track as a dynamic catalyst for creativity, opening new avenues for personalized and expressive music generation. 

\begin{table}[h]
    \caption{Comparison of FAD scores for text-prompted audio generation against target audios (The lower FAD, the better). Audio generated under the text prompts 'Soft Music' and 'Energetic Music' is compared against an evaluation set from the Audiostock dataset labeled with corresponding tags. This comparison demonstrates how text prompts influence the model’s audio generation toward desired perceptual qualities. 
    }
    \centering
    \resizebox{0.5\textwidth}{!}{
    \begin{tabular}{lcc}
    \toprule
    \textbf{Text Prompt} & \multicolumn{2}{c}{\textbf{Target Category}} \\
    \cmidrule{2-3}
     & Soft Music & Energetic Music \\
    \midrule
    Soft Music     & 4.32 & 6.50 \\
    Energetic Music & 6.82 & 3.99 \\
    \bottomrule
    \end{tabular}
    }

    \label{table:Text_Prompts}
    % \vspace{-10pt}
\end{table}

\subsection{Text Conditional Generation}

To elucidate the impact of text prompts on the generation capabilities of our model, we employed the MT-MusicLDM conditioned through CLAP's text encoder branch and used the Audiostock~\cite{chen2023musicldm} dataset as a validation set. We searched for tag words "energetic" and "soft" within the dataset, creating subsets of corresponding audio files. Then, we generated audio files conditioning our model with text prompts "soft music" and "energetic music" through CLAP text encoder, with CFG weight \( w =2.0\). We calculated FAD score across generated and target audios. Table~\ref{table:Text_Prompts} presents the results of this experiment. Notably, the cross-prompt and target folders yielded higher FAD scores, suggesting that the model successfully follows the text prompts. It is worth mentioning that the model was neither trained nor fine-tuned on text, nor on the Audiostock dataset. These results underscore the potential of the model being effectively conditioned on text prompts to influence the perceptual quality of generated music, aligning with our statement that the model represents a step forward towards a versatile audio model.

% \vspace{0.5cm}

\begin{table*}[ht]
    \caption{FAD Scores for instrument stems (B: Bass, D: Drums, G: Guitar, P: Piano) and their combinations in arrangement generation tasks on the Slakh2100 dataset (The lower FAD, the better). The performance of our model is compared to the MSDM baseline across various configurations, using two different FAD calculation protocols— (upper) one for mixtures of generated and provided stems against target mixtures, and (lower) one for generated stems against target stems. }
    \centering
    \resizebox{\textwidth}{!}{
        \begin{tabular}{l|l|cccccccccccccc}
        \toprule
        % \multicolumn{16}{c}{FAD Score $\downarrow$} \\
        % \midrule
        Model & Protocol & B & D & G & P & BD & BG & BP & DG & DP & GP & BDG & BDP & BGP & DGP \\
        \midrule
        MSDM (Baseline) & FAD & 0.45 & 1.09 & \textbf{0.11} & 0.76 & 2.09 & 1.00 & 2.32 & 1.45 & 1.82 & 1.65 & 2.93 & 3.30 & 4.90 & 3.10 \\
        MT-MusicLDM &  Mixture \cite{donahue2023singsong} & \textbf{0.16} & \textbf{0.79} & 0.35 & \textbf{0.34} & \textbf{0.81} & \textbf{0.54} & \textbf{0.68} & \textbf{1.29} & \textbf{1.03} & \textbf{1.00} & \textbf{1.18} & \textbf{0.89} & \textbf{1.03} & \textbf{1.42} \\
        \midrule
        MSDM (Baseline) & FAD  & 6.88 & 5.48 & 4.25 & 6.45 & 4.47 & 6.69 & 6.16 & 4.24 & 4.76 & 6.80 & 4.16 & 4.06 & 5.80 & 4.55 \\
        MT-MusicLDM & Standard & \textbf{0.76} & \textbf{1.07} & \textbf{2.76} & \textbf{1.80} & \textbf{0.89} & \textbf{1.82} & \textbf{1.67} & \textbf{1.40} & \textbf{1.18} & \textbf{2.53} & \textbf{1.30} & \textbf{1.01} & \textbf{2.11} & \textbf{1.42} \\
        \bottomrule
        \end{tabular}
    }

    \label{table:arrange_generation}
    \vspace{-5pt}
\end{table*}

\subsection{Arrangement Generation}
In our arrangement generation experiment, we provided the model with a subset of stems and tasked it with generating the remaining stems unconditionally. We conducted 14 distinct experiments, each focused on generating a specific stem, starting from a singles and expanding to include all possible combinations of them. To assess the model's performance, we calculated the FAD scores for each combination by comparing the generated stems to their corresponding targets from the test dataset.

We chose MSDM as our baseline for the arrangement generation task, as it is the only other open source model known to us capable of generating arrangements in a similar manner. For direct comparisons, we adopted the performance evaluation approach used by MSDM, which in turn utilized a generalized version of the FAD protocol from~\cite{donahue2023singsong}, designed for arrangement generation involving multiple tracks. According to this protocol, generated tracks are mixed with existing originals, and the FAD score is calculated for the resultant mixtures, providing a robust measure of the model’s performance in producing coherent total audio outputs. Additionally, to gain a comprehensive picture, we utilized a publicly available implementation of MSDM
%~\footnote{\href{https://github.com/gladia-research-group/multi-source-diffusion-models}{https://github.com/gladia-research-group/multi-source-diffusion-models}} 
to generate arrangements for all combinations for direct comparison of FAD scores between solely generated stems rather than just mixtures. We pursued this approach because we believe that mixing with given stems can mask a model's performance details, thus not allowing for a detailed analysis of the model's capacity to generate each stem subset.

In the Table~\ref{table:arrange_generation}, we report the FAD score for all instrument stems  (B: Bass, D: Drums, G: Guitar, P: Piano) and their combinations. The upper two rows of table present a comparison of FAD scores for mixes following the protocol described above, while the lower two rows detail the FAD scores between generated tracks and their targets. Our model significantly outperforms MSDM in every combination except for guitar stem generation. By analyzing the values in the bottom row and reflecting on our listening experiences during experiments, we noticed that our model demonstrates notably stronger performance on drums and bass compared to guitar and piano outputs, which are occasionally slightly inferior, as evidenced by the scores (0.76 and 1.07 versus 2.76 and 1.80), and at times exhibit similarities with each other. This observation is further supported, notably by the GP combination, which refers to guitar and piano pair generation and registers as the highest FAD score among all combination categories. Additionally, we observed that when drums are not provided and the model lacks clear rhythmic cues, it often struggles to maintain rhythmic coherence of generated tracks with the given ones. This limitation is reflected by slightly higher FAD scores for the combinations where drums are not included in provided stems. Addressing these limitations and achieving balanced performance across all stems poses an interesting challenge for future research.

\section{Conclusion and Future Work}

We proposed the MT-MusicLDM model, a versatile framework designed to empower creators to generate and compose music in a variety of modes. This includes total track-by-track generation, conditioning generation with reference music tracks or textual inputs, and creating arrangements using any combination of given and generated instrument tracks. Our experiments and evaluations demonstrate that our model produces high-quality sounds across these generative tasks, achieving musical coherence and significantly outperforming the baseline. This work opens several promising avenues for future research.

Although our generative modeling MT-MusicLDM model has shown significant results, limitations remain. These limitations stem from the fact that our model relies on pre-trained components from MusicLDM, such as the VAE and HiFi-GAN vocoder, which are not fine-tuned and specialized in processing individual stems but were trained on music mixes. Additionally, the use of the Mel-spectrogram domain further limits our model's capacity to yield high-fidelity state-of-the-art audio compared to commercial counterparts. Another source of limitation is the choice of dataset, as Slakh2100 is very small considering the data-extensive nature of diffusion models and the task of audio generation. Furthermore, the Slakh dataset doesn't contain any tags or textual descriptions for genre or any other contextual information about its musical pieces.

Looking ahead, we aim to enhance the musical and rhythmic coherence of our model and increase its versatility by expanding the list of available instruments, potentially allowing for user-specified combinations. We intend to extend our investigations by moving away from the Mel-spectrogram domain and possibly shifting to higher sample rates. We plan to incorporate more state-of-the-art VAEs and move to larger datasets beyond Slakh, including large text-to-music datasets, as demonstrated in \cite{donahue2023singsong}. Additionally, we will explore incorporating different methods of conditioning for source separation to achieve truly versatile music generation—a comprehensive music generation framework that supports a wide range of creative expressions, including generation, arrangement, and separation.

In the future, we also see potential for MT-MusicLDM to become an interactive tool for both amateur musicians and educational purposes. For example, it could be developed as a web-based system where users can generate full musical pieces or arrangements based on selected tracks or styles, helping them explore different musical directions. The model could also offer the ability to generate similar tracks or continue existing ones, making it an intuitive tool for creativity and learning.

\section{Acknowledgments}
We thank the Institute for Research and Coordination in Acoustics and Music (IRCAM) and Project REACH: Raising Co-creativity in Cyber-Human Musicianship for their support. This project received support and resources in the form of computational power from the European Research Council (ERC REACH) under the European Union’s Horizon 2020 research and innovation programme (Grant Agreement 883313).

\bibliographystyle{splncs04}
\bibliography{multi_track}

\begin{thebibliography}{10}
\providecommand{\url}[1]{\texttt{#1}}
\providecommand{\urlprefix}{URL }
\providecommand{\doi}[1]{https://doi.org/#1}

\bibitem{borsos2023audiolm}
Borsos, Z., Marinier, R., Vincent, D., Kharitonov, E., Pietquin, O., Sharifi, M., Roblek, D., Teboul, O., Grangier, D., Tagliasacchi, M., Zeghidour, N.: Audiolm: {A} language modeling approach to audio generation. {IEEE} {ACM} Trans. Audio Speech Lang. Process.  \textbf{31},  2523--2533 (2023)

\bibitem{chen2023musicldm}
Chen, K., Wu, Y., Liu, H., Nezhurina, M., Berg-Kirkpatrick, T., Dubnov, S.: Musicldm: Enhancing novelty in text-to-music generation using beat-synchronous mixup strategies. arXiv:2308.01546 (2023)

\bibitem{Deruty2022}
Deruty, E., Grachten, M., Lattner, S., Nistal, J., Aouameur, C.: On the development and practice of ai technology for contemporary popular music production. Transactions of the International Society for Music Information Retrieval  (Feb 2022)

\bibitem{donahue2023singsong}
Donahue, C., Caillon, A., Roberts, A., Manilow, E., Esling, P., Agostinelli, A., Verzetti, M., Simon, I., Pietquin, O., Zeghidour, N., Engel, J.: Singsong: Generating musical accompaniments from singing. arXiv:2301.12662 (2023)

\bibitem{MusGan2018}
Dong, H., Hsiao, W., Yang, L., Yang, Y.: Musegan: Multi-track sequential generative adversarial networks for symbolic music generation and accompaniment. In: Proceedings of the Thirty-Second {AAAI} Conference on Artificial Intelligence, (AAAI-18), the 30th innovative Applications of Artificial Intelligence (IAAI-18), and the 8th {AAAI} Symposium on Educational Advances in Artificial Intelligence (EAAI-18), New Orleans, Louisiana, USA, February 2-7, 2018. pp. 34--41. {AAAI} Press (2018)

\bibitem{MMM2020}
Ens, J., Pasquier, P.: Mmm : Exploring conditional multi-track music generation with the transformer. arXiv:2008.06048 (2020)

\bibitem{Forsgren_Martiros_2022}
Forsgren, S., Martiros, H.: Riffusion - stable diffusion for real-time music generation  (2022), \url{https://riffusion.com/about}

\bibitem{grachten2020bassnet}
Grachten, M., Lattner, S., Deruty, E.: {BassNet}: A variational gated autoencoder for conditional generation of bass guitar tracks with learned interactive control. Applied Sciences  (2020)

\bibitem{ho2020denoising}
Ho, J., Jain, A., Abbeel, P.: Denoising diffusion probabilistic models. In: NeurIPS. vol.~33, pp. 6840--6851 (2020)

\bibitem{ho2022classifierfree}
Ho, J., Salimans, T.: Classifier-free diffusion guidance. In: NeurIPS Workshop on Deep Generative Models and Downstream Applications (2021)

\bibitem{huang2023noise2music}
Huang, Q., Park, D.S., Wang, T., Denk, T.I., Ly, A., Chen, N., Zhang, Z., Zhang, Z., Yu, J., Frank, C., Engel, J., Le, Q.V., Chan, W., Chen, Z., Han, W.: Noise2music: Text-conditioned music generation with diffusion models. arXiv:2302.03917 (2023)

\bibitem{huang2023makeanaudio}
Huang, R., Huang, J., Yang, D., Ren, Y., Liu, L., Li, M., Ye, Z., Liu, J., Yin, X., Zhao, Z.: Make-an-audio: Text-to-audio generation with prompt-enhanced diffusion models. In: ICML. vol.~202, pp. 13916--13932 (2023)

\bibitem{ijcai2022p577}
Huang, R., Lam, M.W.Y., Wang, J., Su, D., Yu, D., Ren, Y., Zhao, Z.: Fastdiff: A fast conditional diffusion model for high-quality speech synthesis. In: IJCAI. pp. 4157--4163 (2022)

\bibitem{MTMG20}
Jin, C., Wang, T., Liu, S., Tie, Y., Li, J., Li, X., Lui, S.: A transformer-based model for multi-track music generation. Int. J. Multim. Data Eng. Manag.  \textbf{11}(3),  36--54 (2020)

\bibitem{karchkhadze2024latent}
Karchkhadze, T., Kavaki, H.S., Izadi, M.R., Irvin, B., Kegler, M., Hertz, A., Zhang, S., Stamenovic, M.: Latent clap loss for better foley sound synthesis. arXiv:2403.12182 (2024)

\bibitem{kilgour2019frechet}
Kilgour, K., Zuluaga, M., Roblek, D., Sharifi, M.: Fr{\'{e}}chet audio distance: {A} reference-free metric for evaluating music enhancement algorithms. In: Interspeech. pp. 2350--2354 (2019)

\bibitem{kingma2022autoencoding}
Kingma, D.P., Welling, M.: Auto-encoding variational bayes. In: Bengio, Y., LeCun, Y. (eds.) ICLR (2014)

\bibitem{kong2020hifi}
Kong, J., Kim, J., Bae, J.: Hifi-gan: Generative adversarial networks for efficient and high fidelity speech synthesis. In: NeurIPS. vol.~33, pp. 17022--17033 (2020)

\bibitem{kong2021diffwave}
Kong, Z., Ping, W., Huang, J., Zhao, K., Catanzaro, B.: Diffwave: {A} versatile diffusion model for audio synthesis. In: ICLR (2021)

\bibitem{lattner2019high}
Lattner, S., Grachten, M.: High-level control of drum track generation using learned patterns of rhythmic interaction. In: WASPAA (2019)

\bibitem{li2023jen}
Li, P., Chen, B., Yao, Y., Wang, Y., Wang, A., Wang, A.: Jen-1: Text-guided universal music generation with omnidirectional diffusion models. arXiv preprint arXiv:2308.04729  (2023)

\bibitem{MIDISandwich19}
Liang, X., Wu, J., Yin, Y.: Midi-sandwich: Multi-model multi-task hierarchical conditional {VAE-GAN} networks for symbolic single-track music generation. Aust. J. Intell. Inf. Process. Syst.  \textbf{15}(2), ~1--9 (2019)

\bibitem{pmlr-v202-liu23f}
Liu, H., Chen, Z., Yuan, Y., Mei, X., Liu, X., Mandic, D., Wang, W., Plumbley, M.D.: {A}udio{LDM}: Text-to-audio generation with latent diffusion models. In: ICML. vol.~202, pp. 21450--21474 (2023)

\bibitem{liu2023audioldm}
Liu, H., Yuan, Y., Liu, X., Mei, X., Kong, Q., Tian, Q., Wang, Y., Wang, W., Wang, Y., Plumbley, M.D.: Audioldm 2: Learning holistic audio generation with self-supervised pretraining. {IEEE} {ACM} Trans. Audio Speech Lang. Process.  \textbf{32},  2871--2883 (2024)

\bibitem{9746901}
Lu, Y.J., Wang, Z.Q., Watanabe, S., Richard, A., Yu, C., Tsao, Y.: Conditional diffusion probabilistic model for speech enhancement. In: ICASSP. pp. 7402--7406 (2022)

\bibitem{luo2023difffoley}
Luo, S., Yan, C., Hu, C., Zhao, H.: Diff-foley: Synchronized video-to-audio synthesis with latent diffusion models. In: NeurIPS (2023)

\bibitem{Slakh}
Manilow, E., Wichern, G., Seetharaman, P., Le~Roux, J.: Cutting music source separation some slakh: A dataset to study the impact of training data quality and quantity. In: WASPAA. pp. 45--49 (2019)

\bibitem{mariani2024multisource}
Mariani, G., Tallini, I., Postolache, E., Mancusi, M., Cosmo, L., Rodol{\`{a}}, E.: Multi-source diffusion models for simultaneous music generation and separation. In: The Twelfth International Conference on Learning Representations, {ICLR} 2024, Vienna, Austria, May 7-11, 2024. OpenReview.net (2024)

\bibitem{melechovsky2024mustango}
Melechovsky, J., Guo, Z., Ghosal, D., Majumder, N., Herremans, D., Poria, S.: Mustango: Toward controllable text-to-music generation. In: NAACL (2024)

\bibitem{nistal2024diffariff}
Nistal, J., Pasini, M., Aouameur, C., Grachten, M., Lattner, S.: Diff-a-riff: Musical accompaniment co-creation via latent diffusion models. arXiv:2406.08384 (2024)

\bibitem{novack2024ditto}
Novack, Z., McAuley, J.J., Berg{-}Kirkpatrick, T., Bryan, N.J.: {DITTO:} diffusion inference-time t-optimization for music generation. In: Forty-first International Conference on Machine Learning, {ICML} 2024, Vienna, Austria, July 21-27, 2024. OpenReview.net (2024)

\bibitem{Paiement2006ProbabilisticMH}
Paiement, J.F., Eck, D., Bengio, S.: Probabilistic melodic harmonization. In: Canadian Conference on AI (2006)

\bibitem{stemgen2024}
Parker, J.D., Spijkervet, J., Kosta, K., Yesiler, F., Kuznetsov, B., Wang, J.C., Avent, M., Chen, J., Le, D.: Stemgen: A music generation model that listens. In: ICASSP 2024 - 2024 IEEE International Conference on Acoustics, Speech and Signal Processing (ICASSP). pp. 1116--1120 (2024)

\bibitem{pasini2024bass}
Pasini, M., Grachten, M., Lattner, S.: Bass accompaniment generation via latent diffusion. arXiv:2402.01412 (2024)

\bibitem{RonnebergerFB15}
Ronneberger, O., Fischer, P., Brox, T.: U-net: Convolutional networks for biomedical image segmentation. In: MICCAI. vol.~9351, pp. 234--241 (2015)

\bibitem{schneider2023mousai}
Schneider, F., Kamal, O., Jin, Z., Schölkopf, B.: Mo\^usai: Text-to-music generation with long-context latent diffusion. arXiv:2301.11757 (2023)

\bibitem{simon2008mysong}
Simon, I., Morris, D., Basu, S.: {MySong}: automatic accompaniment generation for vocal melodies. In: SIGCHI (2008)

\bibitem{Sohl-DicksteinW15}
Sohl{-}Dickstein, J., Weiss, E.A., Maheswaranathan, N., Ganguli, S.: Deep unsupervised learning using nonequilibrium thermodynamics. In: ICML. vol.~37, pp. 2256--2265 (2015)

\bibitem{song2022denoising}
Song, J., Meng, C., Ermon, S.: Denoising diffusion implicit models. In: ICLR (2021)

\bibitem{song2019}
Song, Y., Ermon, S.: Generative modeling by estimating gradients of the data distribution. In: NIPS. vol.~32 (2019)

\bibitem{song2021}
Song, Y., Sohl{-}Dickstein, J., Kingma, D.P., Kumar, A., Ermon, S., Poole, B.: Score-based generative modeling through stochastic differential equations. In: ICLR (2021)

\bibitem{van2017neural}
{van den}~Oord, A., Vinyals, O., et~al.: Neural discrete representation learning. Advances in neural information processing systems  \textbf{30} (2017)

\bibitem{wu2023music}
Wu, S., Donahue, C., Watanabe, S., Bryan, N.J.: Music controlnet: Multiple time-varying controls for music generation. {IEEE} {ACM} Trans. Audio Speech Lang. Process.  \textbf{32},  2692--2703 (2024)

\bibitem{wu2023largescale}
Wu, Y., Chen, K., Zhang, T., Hui, Y., Berg-Kirkpatrick, T., Dubnov, S.: Large-scale contrastive language-audio pretraining with feature fusion and keyword-to-caption augmentation. In: Proc. ICASSP. pp.~1--5 (2023)

\bibitem{xu2024multisourcemusic}
Xu, Z., Dutta, D., Wei, Y.L., Choudhury, R.R.: Multi-source music generation with latent diffusion. arXiv:2409.06190 (2024)

\bibitem{Yang2022}
Yang, D., Yu, J., Wang, H., Wang, W., Weng, C., Zou, Y., Yu, D.: Diffsound: Discrete diffusion model for text-to-sound generation. {IEEE} {ACM} Trans. Audio Speech Lang. Process.  \textbf{31},  1720--1733 (2023)

\bibitem{yao2023jen1}
Yao, Y., Li, P., Chen, B., Wang, A.: Jen-1 composer: A unified framework for high-fidelity multi-track music generation. arXiv:2310.19180 (2023)

\bibitem{Yeh2020}
Yeh, Y.C., Hsiao, W.Y., Fukayama, S., Kitahara, T., Genchel, B., Liu, H.M., Dong, H.W., Chen, Y., Leong, T., Yang, Y.H.: Automatic melody harmonization with triad chords: A comparative study. arXiv:2001.02360  (2020)

\bibitem{yuan2023textdriven}
Yuan, Y., Liu, H., Liu, X., Kang, X., Wu, P., Plumbley, M.D., Wang, W.: Text-driven foley sound generation with latent diffusion model. arXiv:2306.10359 (2023)

\bibitem{zhang2023survey}
Zhang, C., Zhang, C., Zheng, S., Zhang, M., Qamar, M., Bae, S.H., Kweon, I.S.: A survey on audio diffusion models: Text to speech synthesis and enhancement in generative ai. arXiv:2303.13336 (2023)

\bibitem{zhu2023edmsound}
Zhu, G., Wen, Y., Carbonneau, M.A., Duan, Z.: Edmsound: Spectrogram based diffusion models for efficient and high-quality audio synthesis. arXiv:2311.08667 (2023)

\end{thebibliography}

\end{document}